\documentclass[preprint,aps,shortbibliography,natbib=true]{revtex4-2}
\usepackage{graphicx} 
\usepackage{amsmath}
\usepackage{amssymb}
\usepackage{bm}
\usepackage{physics}
\usepackage{hyperref}
\usepackage{color}
\usepackage{soul}

\newcommand{\bs}[1]{\ensuremath{\boldsymbol{#1}}}

\begin{document}
\title{Non-collinear Altermagnetic Phases in the Mott Insulator NiS$_2$}
\author{Mengli Hu$^{1}$, Mikel I. Iraola$^{1}$, Paul McClarty$^{2}$, Jeroen van den Brink$^{1,3,\sharp}$, Maia G. Vergniory$^{4,5,6,\sharp}$
}

\affiliation{
\\$^{1}$Leibniz Institute for Solid State and Materials Research, IFW Dresden, Helmholtzstraße 20, 01069 Dresden, Germany
\\$^{2}$Laboratoire Léon Brillouin, CEA, CNRS, Université Paris-Saclay, CEA-Saclay, 91191 Gif-sur-Yvette, France
\\$^{3}$Würzburg-Dresden Cluster of Excellence Ct.qmat, Technische Universitat Dresden, 01062, Dresden, Germany
\\$^{4}$ Département de Physique et Institut Quantique, Université de Sherbrooke, Sherbrooke,
J1K 2R1 Québec, Canada. 
\\$^{5}$Donostia International Physics Center, 20018 Donostia-San Sebastian, Spain
\\$^{6}$Regroupement Québécois sur les Matériaux de Pointe (RQMP), Quebec H3T 3J7, Canada
\\$^{\sharp}$Corresponding authors: j.van.den.brink@ifw-dresden.de, maia.vergniory@usherbrooke.ca
}

\begin{abstract}

Altermagnets (A$\ell$Ms) constitute a novel family of magnetic materials characterized by the absence of net magnetization and the presence of spin-polarized band structures.
Whereas A$\ell$M phases were initially proposed in collinear structures, the recently discovered noncollinear chiral A$\ell$Ms stand out for their distinct hedgehog spin texture and multifunctionality in spintronics.
In this work, we deepen the characterization of these systems by constructing a Landau theory for noncollinear \emph{achiral} A$\ell$Ms. Furthermore, we demonstrate that the achiral symmetry of the crystal is reflected in the spin texture in reciprocal space, which presents only spatial-even multipoles. These multipoles, distinguished from those in collinear A$\ell$Ms via the high-order secondary order parameters, can couple to many phenomena such as the spin Hall effect and piezomagnetic effect.
To exemplify our theory, we study the noncollinear achiral magnet NiS$_2$ within the framework of altermagnetism, showcasing both spin Hall and piezomagnetic effects in a prototypical correlated Mott insulator that provides an ideal platform to explore the interplay between strong electronic correlations, crystal symmetry, and altermagnetic spin textures.
Interestingly, altermagnetism emerges in two magnetic ordered phases of NiS$_2$ upon lowering the temperature. The non-collinearity strengthens the robustness of A$\ell$M order, as the anti-ferromagnetism induced by the strong correlations will not impose effective time-reversal symmetry as in the collinear case.
Our results suggest non-collinear achiral A$\ell$Ms as a promising platform for spintronics applications due to the potential to achieve various spin textures with different magnetic orders.
\end{abstract}

\maketitle

\section{Introduction}


Altermagnets (A$\ell$Ms) are compensated magnetic phases that share features of both ferromagnets and antiferromagnets \cite{AM_1,AM_2}. 
Proposed first in collinear antiferromagnetic (AFM) systems, the concept has been extended to non-collinear systems, where the local magnetic moments are not aligned along the same direction, but might form non-collinear patterns such as helical\cite{p-wave_helix1,p-wave_helix2,EuIn2As2_helical-EE}, chiral\cite{Mirsi_SHE_Edelstein}, coplanar\cite{AM_landau,p-wave}, or skyrmionic\cite{skyrmion-hall,skyrmion-hall_2d,skyrmion-hall_3}. 
The non-collinearity expands the spin degree of freedom from scalar to pseudo-vector~\cite{AM_noncollinear,odd-parity_criteria,ssg_coplanar-even-wave,colour-spintexture-noncollinear}, thus enriching the field of altermagnetism~\cite{AM_review,jungwirth2025symmetrymicroscopyspectroscopysignatures}.
This generalization enlarges the set of material systems that
obey the general rules of altermagnetism: absence of net magnetization and the presence of spin-polarized band structures, even in the absence of spin-orbit coupling (SOC).


These rules are encoded in the symmetry constraints on A$\ell$Ms and are reflected in the exotic properties of altermagnetic Fermi surfaces 
-- such as the $d$-wave spin polarization and hedgehog spin texture displayed, respectively, by the Fermi surfaces of collinear and non-collinear A$\ell$Ms. Compared to ferromagnets, A$\ell$M Fermi surfaces exhibit strong spin anisotropy~\cite{AM_1,AM_2,p-wave_resistivity_CeNiAsO}, and the non-collinear A$\ell$Ms greatly extend the catalog of magnetic spin textures. Moreover, unlike ferromagnets, A$\ell$Ms are free of magnetic stray fields due to their vanishing magnetization.
Altogether, A$\ell$Ms constitute a promising avenue to combine the respective benefits of ferro- and antiferromagnets for spintronics applications based on their unique magnetic and electronic properties~\cite{AM_1,AM_2,AFMSpintronics1,AFMSpintronics2,p-wave_photocurrent,SOT_noncollinear}.


The recent extension of Landau theory to altermagnets~\cite{AM_landau,AM_landau_collinear,Mirsi_SHE_Edelstein} and the framework of spin space groups~\cite{SSG_1,p-wave} have established the general condition of a spin-split band structure: for collinear A$\ell$M, the magnetic crystal must lack time-reversal combined with inversion ($\cal{PT}$), whereas for non-collinear magnetic ordering, the spin space point group must also not be dihedral.
Furthermore, pseudo-primary order parameters applicable to A$\ell$Ms have been introduced within the framework of Landau theory, based on the N\'{e}el order parameter defined in real space. These order parameters couple to many essential physical observables, and provide information about the underlying spin texture~\cite{AM_1,AM_2,AM_band2,AM_band1,CSVL_1,CSVL_2,multipole_2}. For instance, non-collinear chiral A$\ell$Ms stand out for exhibiting a variety of distinct odd-parity order parameters and multiplets~\cite{Mirsi_SHE_Edelstein}, which allows for the coexistence of the spin-Hall and Edelstein effects in the same system. 

\begin{figure}[h!]
    \centering
    \includegraphics[width=0.5\linewidth]{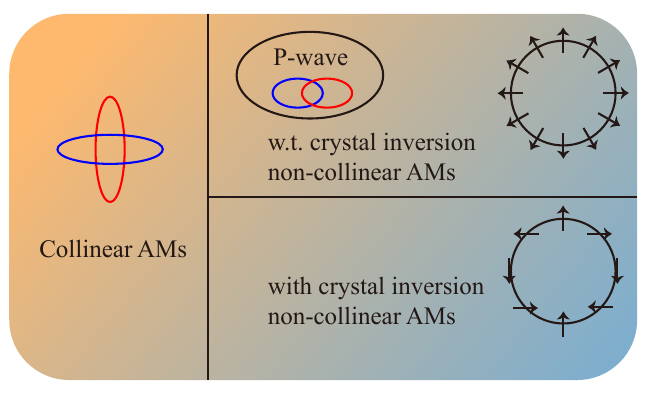}
    \caption{Classification diagram of altermagnetic systems. The horizontal and vertical axes denote (non-)collinearity and (a)chirality, respectively. For each class, the spin texture in reciprocal space is shown schematically with colors and arrows.}
    \label{fig:classification}
\end{figure}

However, the group of non-collinear altermagnets investigated so far concern A$\ell$Ms without crystal inversion.
This includes materials with coplanar helical magnetic order but also more involved chiral magnetic structures that explore all three dimensions of spin space.
For now, many non-collinear odd-parity A$\ell$Ms have been idenitified numerically~\cite{p-wave,p-wave_EE,odd-parity_criteria,odd-parity_exchange,odd-parity_orbital,odd-wave_CrSe,non-collinear_sc-diode-Mn3Pt,non-collinear_NiSi,colour-spintexture-noncollinear,EuIn2As2_helical-EE,p-wave_photocurrent} and experimentally~\cite{p-wave_helix1,p-wave_helix2,p-wave_resistivity_CeNiAsO,odd-parity_EuAuSb,non-collinear_mn-based} in studying the group theory classification~\cite{odd-parity_criteria,colour-spintexture-noncollinear}, electrical switching~\cite{p-wave_helix1}, transport properties~\cite{p-wave_EE,EuIn2As2_helical-EE,p-wave_modal-transport,p-wave_photocurrent,p-wave_resistivity_CeNiAsO,non-collinear_NiSi,minimalmodels2024,high-order_spincurrent}, superconductivity~\cite{non-collinear_sc-diode-Mn3Pt,p-wave_sc1,p-wave_sc2}, impurity~\cite{p-wave_impurity} in A$\ell$Ms.
Here we point out there is a natural third group in this classification: inversion symmetric non-collinear altermagnets as shown in Fig.~\ref{fig:classification}. 
Inversion symmetry is not broken in these systems and they are therefore achiral, but as we will elaborate, they still exhibit very distinct A$\ell$M characteristics in for instance their spin transport and piezomagnetic properties. 
It is natural to include the inversion in the on-site symmetry of magnetic atoms for searching potential altermagnet candidates, as $\cal{PT}$ will be broken and inversion preserved upon magnetic ordering. 

To specify and detail the properties of achiral noncollinear A$\ell$Ms, we use in this work the pyrite-type compound NiS$_2$ to exemplify this class of systems.
This compound has a rich history and generated significant interest due to the diverse noncollinear phases it exhibits over different temperature ranges. The intriguing properties of its electronic structure also garnered considerable attention recently. It has been widely considered as a platform to study metal-to-insulator transitions driven by pressure\cite{Schuster2012, Park2024,  Friedemann2016, Tao2024}, gating~\cite{Day-Roberts2023} and chemical substitution of S~\cite{Yao1996, Kunes2010, Xu2014, Moon2015, Han2018, Jang2021}. Whereas the material  is insulating in the bulk~\cite{Fujimori1996, Matuura1998, Iwaya2004, Kunes2010, Tao2024}, transport measurements have reported the transfer of charge at the surface\cite{Thio1994, Sarma2003, Rao2011, Clak2016, El-khatib2021, El-khatib2023}. In addition, metallic hinges have been observed via scanning tunneling measurements at the step-edges of the material~\cite{Yasui2024}.  
Furthermore, a recent work has classified NiS$_2$ as an obstructed atomic limit, and proposed the obstructed Wannier charges steming from the S-S bonds as the origin of both the 1D metallic hinges and the surface transport~\cite{Iraola2025}. Therefore, NiS$_2$ constitutes a promising material as it exhibits a rich variety of achiral noncollinear phases, strongly-interacting physics, exotic surface properties and a robust Wannier obstruction.
%
Here, in Section \ref{sec_LT}, we elaborate its A$\ell$M Landau theory and construct it for high-T NiS$_2$ with electronic analysis of two magnetic phases. 
In Section \ref{sec_SHE_PZM}, the results of spin Hall effect and piezomagnetism are presented alongside with the evolutions of spin texture and Fermi surface.
Finally, we conclude with a summary and outlook in Section \ref{sec_con}.



\section{Landau theory of non-collinear achiral A$\ell$M}\label{sec_LT}

To elaborate on the properties of non-collinear achiral A$\ell$Ms, it is useful to briefly review their counterparts: collinear achiral and non-collinear chiral A$\ell$Ms. 
A$\ell$M was first proposed in collinear achiral systems~\cite{AM_1,AM_2}, where inversion or mirror symmetries are present. 
The Landau theory of A$\ell$M systems reads $F = c_2 \boldsymbol{\Psi} \cdot \boldsymbol{\Psi} + c_4 (\boldsymbol{\Psi} \cdot \boldsymbol{\Psi})^2$, with the order parameter $\boldsymbol{\Psi}$ describing the N\'eel order without SOC~\cite{AM_landau,AM_landau_collinear}. The altermagnetic constraints are naturally imposed via the irreducible representation (IR) of $\boldsymbol{\Psi}$, which belongs to a non-trivial one-dimensional IR $\Gamma$ of the paramagnetic phase (PM) point group while transforming like a vector in spin space.
It is also possible to define a spatial secondary (pseudo-primary) order parameter $\boldsymbol{O}_{\Gamma} = \int d^3\mathbf{r} [ r_{\mu_1} \ldots r_{\mu_p}] \boldsymbol{m}(\mathbf{r})$ which couples linearly to $\boldsymbol{\Psi}$ via a term $\lambda \boldsymbol{\Psi} \cdot \boldsymbol{O}$. This secondary order parameter transforms according to the same IR as $\boldsymbol{\Psi}$, and contains information about the spin density $\boldsymbol{m}(\mathbf{r})$ and the spatial coordinates $[ r_{\mu_1} \ldots r_{\mu_p}]$.
Furthermore, $\boldsymbol{O}$ describes the spin texture in reciprocal space ($\mathbf{r} \rightarrow \mathbf{k}$) -- in fact, it describes the multipolar order parameters such as the magnetic octupole found in rutiles. Since inversion symmetry cannot connect two magnetic sublattices in collinear A$\ell$Ms -- i.\,e. $D_{\Gamma}(I) = 1$, where $D_\Gamma$ denotes the matrices of the IR $\Gamma$ -- $\boldsymbol{O}$ cannot account for spatially-odd multipoles ($D_{\Gamma}(I) \neq 1$). Within the framework of spin-space groups (SSG), this unfeasibility of odd-parity multipoles in collinear A$\ell$Ms can be derived based on the presence of the symmetry $C_{2\bot}T$ acting only on spin space (where $C_{2\bot}$ denotes the 2-fold rotation with respect to the axis perpendicular to the spin), which maps $\boldsymbol{k}$ to $\boldsymbol{-k}$ and keeps the spin invariant.

Non-collinear arrangements of moments and chirality in general allow for spatial-odd multipoles and non-collinear spin textures in A$\ell$Ms (see Fig.~\ref{fig:classification}), such as the hedgehog texture found in MnIr$_3$Si~\cite{Mirsi_SHE_Edelstein}. 
The complex non-collinear structure also induces high-dimensional IR multipoles, such as in MnIr$_3$Si: $\boldsymbol{O}^{\alpha}_{T} = \int d^3\mathbf{r} \boldsymbol{r}_{\alpha} \boldsymbol{s}(\mathbf{r}), \alpha=1,2,3$. We also note that odd-parity spin textures may arise in coplanar magnets in non-centrosymmetric crystals and in the presence of a $[C_{2\perp}||\bf{t}]$ symmetry that enforces compensation of the in-plane moments. The symmetries allow an out-of-plane spin texture of odd parity including p-wave textures~\cite{p-wave,p-wave_helix1,p-wave_helix2,Yu_oddparity_2025,lee_iaAFM_2025}.  
Here, we build the Landau theory for a non-collinear achiral A$\ell$M without imposing strict constraints on the magnetic structure. 
To give this a practical context, we do so by considering NiS$_2$, which hosts two distinct non-collinear centrosymmetric phases~\cite{Iraola2025}, using an approach and methodology that may be applied to other non-collinear achiral A$\ell$Ms in a straightforward manner.

\begin{figure}[htpb]
    \centering
    \includegraphics[width=1.0\linewidth]{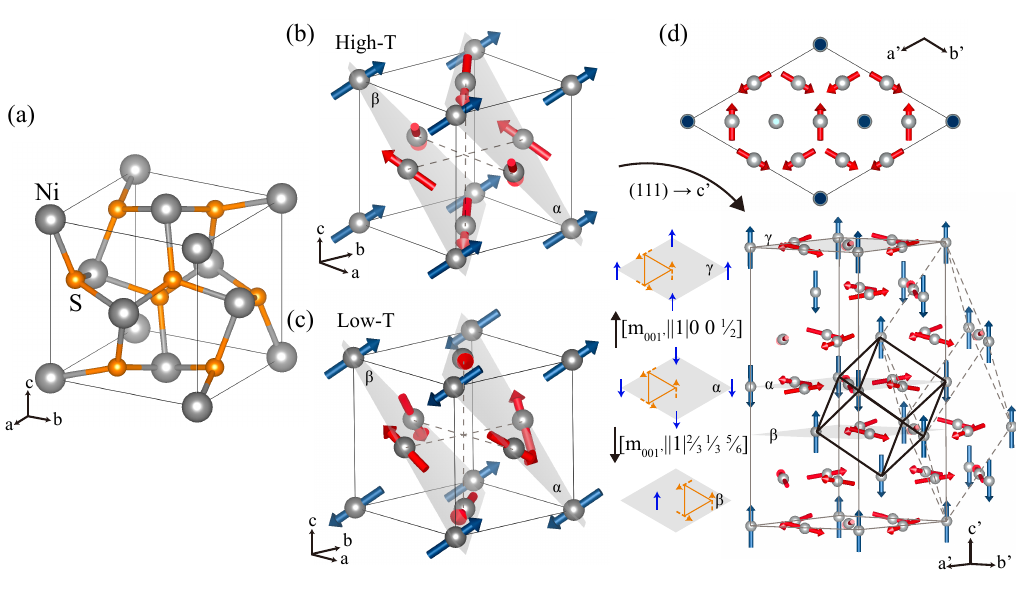}
    \caption{NiS$_2$ crystal structure (a) and magnetic structure with only Ni atoms shown from high- (b) and low- (c,d) temperature (high-T, low-T) phases. The magnetic phase transition happens twice, upon lowering the temperature, from paramagnetic to high-T and high-T to low-T phase. Ni atoms lying at light grey planes ($\alpha$, $\beta$, and $\gamma$), perpendicular to the cubic (111) direction, are equivalent in high-T phase and inequivalent between $\alpha$ and $\beta$/$\gamma$ planes.
    The standard unit cell of low-T phase is shown in (d) with top view at the upper panel.
    The lower panel in (d) shows the relation between different unit cells: the black solid lines and grey dashed lines denote the primitive cells of high-T and low-T phases, respectively.}
    \label{fig:NiS2_crystal}
\end{figure}

\subsection{Characterization of the Two Magnetic Phases in NiS$_2$}

NiS$_2$ crystallizes in a cubic structure with the space group $Pa\bar 3.1'$ (No.205.34), similar to the pyrite compound FeS$_2$.  In this paramagnetic configuration, Ni atoms are located at the Wyckoff position (PS) $4a$, whereas S atoms sit at WP $8c$.
Upon lowering the temperature below $39$ K, the system undergoes a transition into a magnetic phase with vanishing total magnetization, dubbed the high-T phase. The high-T phase is similar to that of the ordered phase of MnTe$_2$ \cite{Zhu2024} that is known to exhibit a complex spin texture in the electronic bands.
Upon further cooling, below $30$ K, magnetic moments realign forming a weak-ferromagnetic structure dubbed the low-T phase.
Within the framework of magnetic space groups, the symmetry of the high-T phase is described by the Fedorov group P$a\bar3$ (No. 205.33). 
The characterization of the low-T phase has remained a subject of debate despite the intensive efforts devoted to it~\cite{Hastings70, Nishihara75, Nagata76, Kikuchi78, Higo2015, Yano16}.
Recently, Iraola \textit{et al.} corroborated $R\bar 3$ (No. 148.17) as the magnetic space group for the low-T phase on the basis of high-quality neutron scattering data, and advanced an appropriate model for its magnetic structure~\cite{Iraola2025}.


To explore potential altermagnetic characteristics of these magnetic phases, we first
illustrate their local magnetic moments using the same crystal unit cell as shown in Fig.~\ref{fig:NiS2_crystal} (a-c). Before diving into the details, we note that the low-T phase doubles the unit cell via the propagation vector $R:(\frac{1}{2},\frac{1}{2},\frac{1}{2})$, and hosts hexagonal standard and rhombohedral primitive unit cell as shown in Fig.~\ref{fig:NiS2_crystal}(d). Fig.~\ref{fig:NiS2_crystal}(d). Both of these lattices host a Cartesian coordinate system with the $z$ axis pointing to the high-T cubic (111) direction. The unit vector relations between the low-T and high-T phases are given by:

\begin{equation}
    (\bs{a}' \ \bs{b}' \ \bs{c}') 
    = 
    (\bs{a} \ \bs{b} \ \bs{c})  
    \begin{bmatrix} 1 & 0 & 2 \\ 0 & 1 & 2\\ -1 & -1 & 2 \\ \end{bmatrix},
     (\bs{a}'^p \ \bs{b}'^p \ \bs{c}'^p) 
    = 
    (\bs{a} \ \bs{b} \ \bs{c})  
    \begin{bmatrix} 1 & 1 & 0 \\ 1 & 0 & 1\\ 0 & 1 & 1 \\ \end{bmatrix},
\end{equation}
where $(\mathbf{a},\mathbf{b},\mathbf{c})$, $(\mathbf{a}',\mathbf{b}',\mathbf{c}')$, and $(\mathbf{a}'^p,\mathbf{b}'^p,\mathbf{c}'^p)$ denote the primitive vectors of the high-T phase, and the conventional vectors and primitive vectors of the low-T phase, respectively; these three cells are denoted in Fig.~\ref{fig:NiS2_crystal}(d) with solid black, grey, and dashed grey lines. 

The magnetic phase transitions occur only with the local magnetic moments ordering/re-ordering, while the crystal and atomic structure remain unchanged. 
One can track the changes in the Ni moment orientations within light-grey planes ($\alpha$, $\beta$) in Fig.~\ref{fig:NiS2_crystal} (b) and (c), and each plane contains a set of Ni: $4a$. The first magnetic ordering happens, where local moments follow the crystal symmetry: $\mathbf{m}_{\text{Ni}_1} = \{2_{001}|\frac{1}{2}0\frac{1}{2}\}\mathbf{m}_{\text{Ni}_2} = \{2_{010}|0\frac{1}{2}\frac{1}{2}\}\mathbf{m}_{\text{Ni}_3}= \{2_{100}|\frac{1}{2}\frac{1}{2}0\}\mathbf{m}_{\text{Ni}_4}$ with $\mathbf{r}_{\text{Ni}_1} = (0,0,0)$ and $\mathbf{m}_{\text{Ni}_1} =(m,m,m), m = 0.569$. The crystal-symmetry-preserving transition leads to the isomorphic relation \cite{SSG_irreps_1,SSG_irreps_2} between the magnetic space group (MSG): $Pa\overline{3}$ (\# 205.33), the spin space group (SSG): $P^{2_{100}}a^{3^1_{111}}-3$ (\# 205.1.12.1), and SG: $Pa\bar 3$ (\# 205).
When lowering the temperature, the magnetic re-ordering occurs and manifests as the local magnetic components perpendicular to $\alpha$/$\beta$. The magnetic moments within $\alpha$ all tilt towards the (111) direction, while the magnetic moments within $\beta$ tilt towards the ($\overline{1}\overline{1}\overline{1}$) direction with the same magnitude.
This re-ordering breaks most of the crystal symmetry and introduces a pure spin mirror symmetry: $\{m_{001}||1|\frac{2}{3} \frac{1}{3} \frac{5}{6}\}$ in the SSG due to the same tilting values from $\alpha$ and $\beta$.
With the only preserved generator $\{-3_{111}|0 0 0\}$, the (111) direction in cubic lattice becomes the main axis in the low-T standard unit cell ($\mathbf{c}'$), and we will use these two different coordinate systems for the high-T and low-T phases in the following analysis.
The unique symmetry: $\{m_{001}||1|\frac{2}{3} \frac{1}{3} \frac{5}{6}\}$ in SSG gives rise to different group structure and IRs between MSG: $R\overline{3}$ (\# 148.17) and SSG: $R^{3^1_{001}}-3|(1,1,m_{001};m_{001},1)$ (\# 148.2.3.2)\cite{SSG_1,SSG_magnon} at low-T phase.\footnote{The international notation is adapted to represent the SSG symmetries. The SSG of high-T phase is a t-type with representative symmetries: $\{2_{100}||m_{100}|\frac{1}{2}\frac{1}{2}0\}$ and $\{3^1_{111}||-3_{111}|0\}$;
the SSG of low-T phase is a g-type with representative symmetries: $\{3^1_{001}||-3_{001}|0\}$, and $\{1||1|1\ 0\ 0\}$, $\{1||1|0\ 1\ 0\}$, $\{m_{001}||1|0\ 0\ 1\}$, $\{m_{001}||1|\frac{2}{3}\ \frac{1}{3}\ \frac{1}{3}\}$, $\{1||1|\frac{1}{3}\ \frac{2}{3}\ \frac{2}{3}\}$.}

\subsection{Landau theory of NiS\texorpdfstring{$_2$}{2} high-T phase}

The first magnetic phase transition occurs from PM to high-T phase with the propagation vector at $\Gamma=(0,0,0)$. Hence, we use the IRs at $\Gamma$ of the PM phase to decompose the four-dimensional reducible representation of Ni atom positions as: $A_g\oplus T_g$. It is noted that both of these IRs are inversion even.

The $A_g$, which is the trivial representation, only describes the total magnetization per unit cell as: $\mathbf{M} = \sum_{a=1}^4 \mathbf{m}_{a}$ instead of the altermagnetic order. By projecting to $T_g$, the basis functions can be: $\mathbf{\Phi}^{\alpha} = \sum_{a=1} ^{4} \phi_{T_g,a}^{\alpha}\mathbf{m}_a$, with $\phi_{T_g}^1 = (a,a,-a,-a)$; $\phi_{T_g}^2 = (a,-a,a,-a)$, $\phi_{T_g}^3 = (a,-a,-a,a)$, where $a$ is a parameter indicating the relation between $\phi_{T_g}$. Based on the primary order parameter belonging to $T_g$, the Landau theory can be written as:
$$
F = c_2 \mathbf{\Phi}^{\alpha} \cdot \mathbf{\Phi}_{\alpha} + c_4 (\mathbf{\Phi}^{\alpha} \cdot \mathbf{\Phi}_{\alpha})^2 + \ldots.
$$

For the inversion-even property of $T_g$, $\mathbf{\Phi}^{\alpha}$ can only couple to the quadrupole/hexadecapole and any spatial-even multipole as the secondary order parameter. The spatial quadrupolar order parameter reads:

$$
\mathbf{L}^{H}_{\alpha} = \int d^3r Q^{H}_{\alpha} \mathbf{m}(\mathbf{r}) \ \ Q^{H}_{\alpha} = (r_xr_y, r_xr_z, r_yr_z).
$$

\begin{figure}[h]
    \centering
    \includegraphics[width=0.8\linewidth]{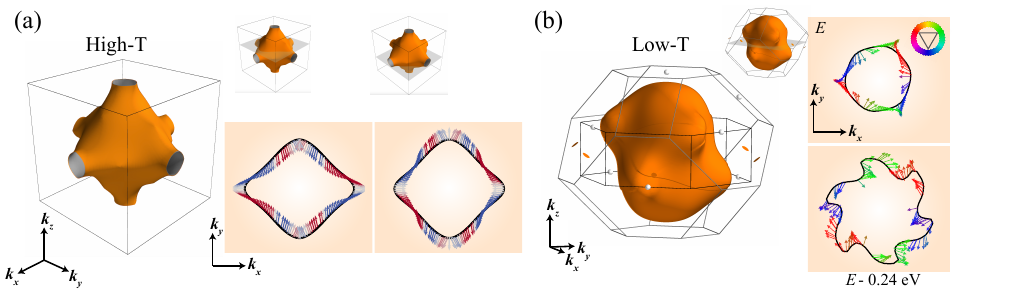}
    \caption{Fermi surfaces and spin textures of NiS$_2$ for the high-T phase (a) and the low-T phase (b). Both phases have finite band gaps. The Fermi surfaces are shown at $E=E_f-0.55\ \mathrm{eV}$ and $E=E_f-0.12\ \mathrm{eV}$ for the high-T and low-T phases, respectively. To illustrate the quadrupolar spin texture, slices at $k_z=\pm0.2\,\pi/c$ and at $k_z=0$ are shown next to each Fermi surface plot. The $s_z$ component in the high-T phase is indicated by the arrow colors in (a). For the low-T phase, all spins are rotated counterclockwise by $\pi/6$, with colors denoting the coplanar spin angles for clarity.}
    \label{fig:spintexture}
\end{figure}

We then compute the electronic and magnetic structure of high-T NiS$_2$. As shown in Fig.~\ref{fig:spintexture} (a), the spin textures at $k_z=\pm 0.2\frac{\pi}{c}$ present a quadrupole distribution without SOC. The component $s_z(\mathbf{k})$, scaled with a blue-to-red colormap, presents a $d$-wave-like spin polarization. However, the two-dimensional spin texture ($\mathbf{s}_{\perp} = (s_x, s_y)$) at $\pm k_z$ planes shows a Rashba-type feature. From the spin texture, we expect that both the extrinsic and intrinsic mechanisms of the spin Hall effect are relevant. For instance, $s_z$ polarized spin current in $x-y$ plane is contributed extrinsically from the electron spin-$z$ component, whose velocity are identical between $\pm k_z$: $v_{\perp}(k_z) = v_{\perp}(-k_z)$. There is another mechanism via the torque under electric field ($E$): $\dot{s_z}\approx (\mathbf{s} \cross \mathbf{s}(E))_z$. Even if the in-plane spin flips sign $\mathbf{s}_{\perp}(k_z) = -\mathbf{s}_{\perp}(-k_z)$, the induced spin currents are the same: $\dot{s_z}(k_z)$ = $\dot{s_z}(-k_z)$. 
Note that in the absence of SOC, the spin-coordinate frame is energetically degenerate for all directions. The analysis of spin currents presented here therefore remains valid under a global spin rotation, with the spin polarization defined with respect to that spin-coordinate frame. Thus both mechanisms indicate a large spin Hall effect in this system.
With the secondary order parameters, we notice that the spin Hall conductivity is time-reversal even and transforms as the IR: $A_g$. In the SOC-free limit, it can couple to the primary order parameter as $T_g \otimes T_g$, which contains the $A_g$ IR. The numerical results based on DFT are presented in the next section.

\subsection{Low-T phase and its spin texture}

The magnetic transition occurs again upon lowering the temperature, along with the re-ordering of local magnetic moments. Within the limit of the absence of SOC, its net-zero magnetization ordering is protected via its SSG: $R^{3^1_{001}}-3|(1,1,m_{001};m_{001},1)$~\cite{SSG_1}.
As this magnetic transition happens from one magnetic ordered phase to another, where the $SU(2)$ spin space is already broken, we analyze the altermagnetic properties, including the spin texture, merely based on the magnetic structure of the low-T phase.

It is noted that all atomic sites remain the same for all phases. Here, we change the coordinate to the conventional unit cell of the low-T phase by rotating the (111) direction to the $c'/z'$ axis, as shown in Fig.~\ref{fig:NiS2_crystal}, for the convenience of the following discussion. 
We notice that the $m_{z}$ component changes sign between atoms, as indicated by the propagation vector $R$ in the high-T phase\cite{Iraola2025}. This change, as discussed in the last section, breaks most of the symmetry in the high-T MSG while introducing a spin space group symmetry: $\{m_{001} || 1 | \frac{2}{3} \frac{1}{3} \frac{5}{6}\}$. Microscopically, this spin space group symmetry and crystal inversion ($\{1||-1\}$) restrict the spin texture to be {\it coplanar} in the whole BZ via the relation: $\{1||-1\}\{m_{001} || 1 | \frac{2}{3} \frac{1}{3} \frac{5}{6}\}s_{z}(\mathbf{k}) = -\{1||-1\}s_{z}(-\mathbf{k}) = -s_{z}(\mathbf{k}) =s_{z}(\mathbf{k})= 0$ \cite{note}. As shown in Fig.~\ref{fig:spintexture} (b), the coplanar spin texture appears distinguished from the high-T spin texture in (a). Besides, the other two generators of the low-T SSG read: $\{1||-1\}$, $\{3_{001}||3_{001}\}$. These, in sum, constrain the spin texture microscopically as~\cite{CSVL_2}:
    
\begin{equation}\label{equ.1}
\begin{aligned}
    \{3_{001}||3_{001}\} \mathbf{s}(\mathbf{k}) &= 3^{-1}_{z}\mathbf{s}(3_{z}\mathbf{k})=\mathbf{s}(\mathbf{k}), \\
    \{1||-1\}\mathbf{s}(\mathbf{k}) & = \mathbf{s}(-\mathbf{k})=\mathbf{s}(\mathbf{k}); \\
    \epsilon(\mathbf{k}) = \epsilon(3_{z}\mathbf{k}), &  \ \ \epsilon(\mathbf{k}) = \epsilon(-\mathbf{k}),
\end{aligned}
\end{equation}
where $\epsilon$ is the eigen-energy.
The constraint of crystal inversion symmetry enforces $\mathbf{s}(-\mathbf{k})=\mathbf{s}(\mathbf{k})$, suggesting that only even-parity multipoles are allowed in the low-T phase. We also observe the spatial-even multipole as the secondary order parameter in the high-T phase, which is due to the fact that crystal inversion symmetry cannot be broken via different magnetic orderings, as the site symmetry group of Ni atoms contains inversion.

The low-T spin texture is shown in Fig.~\ref{fig:spintexture} (b), where the quadrupole appears around the valence band maxima and continuously shows up with a petal-like Fermi surface consistent with the six-fold rotational symmetry at $k_z=0$, as analyzed in Equ.~\ref{equ.1}. 
Both the high-T and low-T phases show even-parity multipoles as expected in theory, while the coplanar spin texture appears in the low-T phase due to the additional pure spin mirror symmetry.  
In terms of piezomagnetism~\cite{AM_landau,PZM_MnTe,CSVL_1}, the additional pure spin mirror symmetry forbids the $m_{z'}$ component through point symmetry breaking. We expect no piezomagnetism of $m_{z'}$ components under any kind of strain. In the next section, we present DFT calculations that indeed show a large in-plane piezomagnetic effect and a zero out-of-plane response under uniaxial strains.

Our spin-texture and Landau-theory analyses align with results from the color-symmetry theory applied to A$\ell$Ms~\cite{Tensorial_AM,colour-spintexture-noncollinear}. The spin textures of the high-T and low-T phases, in tensorial form, are:
\begin{equation}\label{equ.tensorial}
\begin{aligned}
\text{high-T:}\quad s_{x} &= a\,k_y k_z, \quad s_{y} = a\,k_x k_z, \quad s_{z} = a\,k_x k_y, \\
\text{low-T:}\quad s_{x} &= b_1 (k_x^2 - k_y^2) + 2 b_2 k_x k_y + b_3 k_x k_z - b_4 k_y k_z, \\
s_{y} &= -2 b_1 k_x k_y + b_2 (k_x^2 - k_y^2) + b_3 k_y k_z + b_4 k_x k_z, \\
s_{z} &= 0.
\end{aligned}
\end{equation}
Note that the two phases are represented in different coordinate systems. The vanishing out-of-plane spin component in the low-T phase originates from mirror invariance in spin space~\cite{colour-spintexture-noncollinear}: $\mathbf{s}(\mathbf{M}) = m_{z}^{-1}\mathbf{s}(m_{z}\mathbf{M})$, which implies $s_{z} = -s_{z}$ and hence $s_{z}=0$. This mirror symmetry, imposed by the anti-ferromagnetic order in $m_z$, indicates stronger correlations at lower temperatures. Due to the non-collinearity, the low-T phase still shows the spin-split bands with non-zero $m_x-m_y$. Both phases are compatible with the piezomagnetic and magneto-optical Kerr effects~\cite{colour-spintexture-noncollinear}. Their response tensors have the same tensorial form as Eq.~\ref{equ.tensorial}, since they share the same Jahn symbol. From the low-T results, we can extract four quadrupolar order parameters:
\[
\mathbf{L}^{L}_{\alpha} = \int d^3r\, Q^{L}_{\alpha}\,\mathbf{m}(\mathbf{r}),
\]
where, for example $Q^{L}_{\alpha}$ can be,
\begin{gather*}
Q^{L}_{\alpha,1} = (r_x^2 - r_y^2,\ -2r_xr_y),\qquad
Q^{L}_{\alpha,2} = (2r_xr_y,\ r_x^2 - r_y^2),\\
Q^{L}_{\alpha,3} = (r_xr_z,\ r_yr_z),\qquad
Q^{L}_{\alpha,4} = (r_yr_z,\ -r_xr_z).
\end{gather*}
These quadrupolar order parameters are visible in Fig.~\ref{fig:spintexture}(b).

\newpage

\section{Spin Hall and Piezomagnetic effect of NiS\texorpdfstring{$_2$}{2}} \label{sec_SHE_PZM}

As we show via the Landau theory of both high-T and low-T phases, their order parameters are both inversion-even and coupled to the spin Hall and piezomagnetic effects without SOC. 
In other words, these two effects should co-exist in both phases. 
Specifically, the spin Hall effect will appear upon applying an external electric field and is signaled by the transverse spin current. The total spin current is described through the conductivity tensor as: $\mathcal{J}_j^i = \sum_{k} \sigma_{jk}^{i}E_{k}$, where the anti-symmetric part of $\sigma_{jk}$ corresponds to the spin Hall current that is perpendicular to the electric field. The piezomagnetic effect arises from the (uniform) strain field $\epsilon$, which can be classified as uniaxial ($\epsilon_{ii}$) and shear strain ($\epsilon_{ij}$). The net magnetization is induced via strain as: $M^i = \sum_{jk}\Lambda^i_{jk}\epsilon_{jk}$.

The co-existence of piezomagnetic and spin Hall effects lies in the identical response tensors in the absence of SOC. Both of them are rank-three tensors containing two indices that change under symmetry as the normal vectors and one as the axial vectors, with their Jahn symbols being $MV2$\cite{S_tensor}. We already show the non-zero spin texture expressions up to $k_ik_j$ in Equ.~\ref{equ.tensorial} in the same $MV2$. Here we cross-check the symmetry constraints of both high-T and low-T phases, and their response tensors do share the same forms ($\Lambda \sim \sigma$). The piezomagnetic response tensors read: 
\begin{equation}\label{equ.response}
\begin{array}{p{1.5cm}p{0.5cm}p{4cm}p{0.5cm}p{4.6cm}p{0.5cm}p{3.5cm}}
\text{high-T:} & $\Lambda^{x}$ &= $\begin{pmatrix} 0 & 0 & 0 \\ 0 & 0 & \Lambda^{x}_{yz} \\ 0 & \Lambda^{x}_{zy} & 0 \end{pmatrix}$,
& $\Lambda^{y}$ &= $\begin{pmatrix} 0 & 0 & \Lambda^{x}_{zy} \\ 0 & 0 & 0 \\ \Lambda^{x}_{yz} & 0 & 0 \end{pmatrix}$, 
& $\Lambda^{z}$ &= $\begin{pmatrix} 0 & \Lambda^{x}_{yz} & 0 \\ \Lambda^{x}_{zy} & 0 & 0 \\ 0 & 0 & 0 \end{pmatrix}$, \\
\text{low-T:} & $\Lambda^{x}$ &= $\begin{pmatrix}
\Lambda^{x}_{xx} & \Lambda^{x}_{xy} & \Lambda^{x}_{xz} \\
\Lambda^{x}_{xy} & -\Lambda^{x}_{xx} & \Lambda^{x}_{yz} \\
\Lambda^{x}_{xz} & \Lambda^{x}_{yz} & 0
\end{pmatrix}$, 
& $\Lambda^{y}$ &= $\begin{pmatrix}
\Lambda^{x}_{xy} & -\Lambda^{x}_{xx} & -\Lambda^{x}_{yz} \\
-\Lambda^{x}_{xx} & -\Lambda^{x}_{xy} & \Lambda^{x}_{xz} \\
-\Lambda^{x}_{yz} & \Lambda^{x}_{xz} & 0 \end{pmatrix}$, 
& $\Lambda^{z}$ &= 0 .
\end{array}
\end{equation}

\begin{figure}[h!]  
    \centering
    \includegraphics[width=0.9\linewidth]{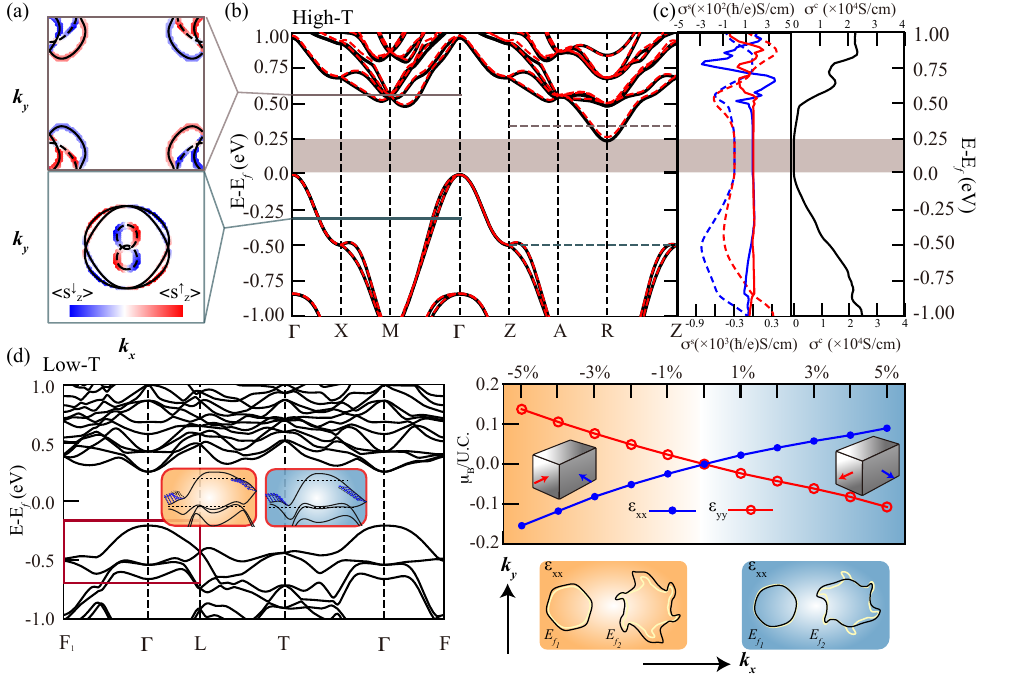}
    \caption{Electronic properties of NiS$_2$ and calculated spin Hall conductivity at high-T phase and piezomagnetism at low-T phase. (a) The spin texture in the $k_z = 0/\pi$ plane taken at $E= E_f - 0.30/-0.50$ (eV) for valence band, and $E= E_f + 0.35/+0.55$ (eV) for conductance band, respectively. (b) The band structure along high symmetry lines without and with SOC in solid black and dashed red lines, respectively. The mBJ approximation is adapted with the mBJ potential CMBJ = 1.15~\cite{mbj_1,mbj_2}. (c) The SHE effect respective to the energy region in (b). The two independent spin conductivity elements $\sigma^{I/II,x}_{yz},\sigma^{I/II,x}_{zy}$ are shown in blue and red in the left panel, and the contributions from $\chi^{I/II}$ are in dashed and solid lines, respectively. The charge conductivity element is shown in the right panel. (d) Band structure and piezomagnetism under uniaxial strain along $x$ and $y$ directions. 
    The strained band structures and corresponding two-dimensional Fermi surfaces evolutions are shown with the same background colors. The right-upper panel shows the relation between strain strength and $M^{y}$.}
    \label{fig:NiS2_band}
\end{figure}

Then we calculate spin Hall conductivity in high-T phase and piezomagnetic effect in low-T phase as examples and analyze their changes through the magnetic phase transition. The linear spin Hall response is calculated from two main contributions~\cite{Zelezny2017,Freimuth2014}: $\mathcal{J}_j^i = \sum_{k} (\sigma_{jk}^{I,i}+\sigma_{jk}^{II,i})E_{k}$, where:
\begin{widetext}
\begin{equation}\label{kubo}
\begin{aligned}
\sigma ^{I,i}_{jk} &= -\frac{e\hbar}{\pi VN}\sum_{\mathbf{k},m,n} \frac{\Gamma^2 \, \mathrm{Re} \left( \mel{n\mathbf{k}}{\frac{1}{2}[\hat{s_i},\hat{v_j}]}{m\mathbf{k}}\mel{m\mathbf{k}}{\hat{v}_k}{n\mathbf{k}} \right)}{[(E_f - \epsilon_{n\mathbf{k}})^2 + \Gamma^2][(E_f - \epsilon_{m\mathbf{k}})^2 + \Gamma^2]} , \\
\sigma ^{I\!I,i}_{jk} & = -\frac{2\hbar e}{VN} \sum_{\mathbf{k},n\neq m}^{\substack{n=\text{occ} \\ m=\text{unocc}}} \frac{ \mathrm{Im} \left( \mel{n\mathbf{k}}{\frac{1}{2}[\hat{s_i},\hat{v_j}]}{m\mathbf{k}}\mel{m\mathbf{k}}{\hat{v}_k }{n\mathbf{k}} \right)}{(\epsilon_{n\mathbf{k}} - \epsilon_{m\mathbf{k}})^2}.
\end{aligned}
\end{equation}
\end{widetext}
Here, $e$ is the elementary charge, $\mathbf{k}$ is the Bloch wave vector, $n, m$ are the band indices, $\epsilon_{n,\mathbf{k}}$ is the eigenvalue, $E_f$ is the Fermi energy, $\hat{\mathbf{v}}$ and $\hat{\mathbf{s}}$ are the velocity and spin operators, respectively. $\Gamma$ is the inverse scattering time, and
$N$ and $V$ are the total number of Bloch waves and the volume of the unit cell.
From band structures in Fig.~\ref{fig:NiS2_band}, the semiconductor properties persist along with the magnetic phase transition, where a fundamental gap appears in both phases with gap values of $\Delta_{high} \approx 0.25$ eV and $\Delta_{low} \approx 0.44$ eV. We noticed that the gap values are 0.80 eV from STM spectra\cite{Iraola2025} and 0.25 $\sim$ 0.29 eV from optical absorption\cite{NiS2_gap} within the temperature region discussed. Here, we set the parameters in calculations based on the experimental local magnetic moments to better investigate the magnetic phase transition\cite{Iraola2025}. 
For the smaller gap of the high-T phase, the independent conductivity elements $\sigma^x_{xy}$ and $\sigma^x_{yz}$ versus energy are calculated by using Equ.~\ref{kubo}. As shown in the right panel of Fig.~\ref{fig:NiS2_band} (b), with electron doping of 0.52 per unit cell ($E_f'=E_f+\Delta_{high}+0.3$ eV) or hole doping of 0.12 per unit cell ($E_f'=E_f-0.3$ eV), the spin current can reach up to $\sim 10^2 (\hbar /e)S/cm$, which is of the same order as non-collinear AFM and A$\ell$M~\cite{SHE_AFM,SHE_cal,SHE_mag_cal,SHE_mag_cal2,SHE_review,SHE_high,Mirsi_SHE_Edelstein}. 
The finite spin Hall effect in the gap is consistent with the obstructed atomic limit reported in a previous study~\cite{Iraola2025}, indicating a surface spin Hall conductivity. In the supplementary information, we also show that surface states exist even without SOC and are, in fact, protected by altermagnetism.
The large spin Hall effect in the high-T phase can be understood via the Rashba-type spin texture in two-dimensional momentum planes as we analyzed in the last section, and every two-dimensional contribution adds up to a final considerable bulk effect.
The spin current conductivity changes quantitatively with temperature and, more importantly, qualitatively, as the spin Hall currents are not necessarily perpendicular to the Hall plane in the low-T phase, reflected as $\sigma^{x}_{xy}\neq0$. The changes in spin current polarization also denote the symmetry breaking from the high-T to low-T phase\cite{Iraola2025}.

The piezomagnetic effect is considered a pivotal phenomenon in altermagnets. The microscopic mechanism can be described via the Fermi surface/valley spin texture evolution under strain\cite{CSVL_2,CSVL_1}. Here, we present that even without SOC, the spin polarization can effectively tilt through exchange coupling in the non-collinear achiral A$\ell$M with a finite gap. 
By applying compressive/tensile uniaxial in-plane strain to low-T NiS$_2$, the three-fold rotational symmetry is broken. As shown in Fig.~\ref{fig:NiS2_band}(d), uniaxial strain along $x$ or $y$ can effectively tune the Fermi surfaces and deform them into symmetry-broken phases. The spins are also rotated under strain as shown in the inset in Fig.~\ref{fig:NiS2_band}(d). Both Fermi surface deformation and spin-tilting suggest strain is an effective tool to induce net magnetization in non-collinear gapped A$\ell$Ms.
The persistence of the zero $M^{z}$ component is preserved by the spin only symmetry, which cannot be broken by any uniaxial strain.
The net magnetization is up to $\sim\pm 0.15 \mu_B$ per unit cell with $\pm$ 5\% strain strength. 
Compared between the value of strain-induced net magnetization: $M^{x}$ and $M^{y}$, our numerical results show that $M^{y}$ is always one order larger than $M^{x}$. One can change its sign by either changing the sign of strain or swapping between uniaxial strain directions of $x$ and $y$ as shown in Fig.~\ref{fig:NiS2_band}(d).
The zero response in the $M^{z}$ component of the low-T phase suggests that neither spin current nor net magnetization will appear under strain or electric field. Furthermore, uniaxial strain cannot induce net magnetization in the low-T phase, as presented in the response tensors. These differences and the polarization change of spin Hall current between the two phases can serve as signatures of the magnetic phase transition.

\section{Discussion and Conclusion}\label{sec_con}

The semiconducting nature of both the high- and low-T magnetic phases of NiS$_2$ has recently been reported~\cite{Iraola2025}. Here, detailed analysis of the magnetic phases, shows that both the high- and low-T phases are altermagnetic, with inversion-even order parameters that couple to spin Hall and piezomagnetic effects even in the absence of spin-orbit coupling. Our first-principles calculations show that both effects are large in magnitude and change quantitatively and qualitatively through the magnetic phase transition. 
Moreover, this is the first established case for piezomagnetism emerging from non-collinear altermagnetic order while the electronic structure stays gapped. 
The different SSG symmetries of the two phases lead to distinct spin textures, non-coplanar in high-T and coplanar in low-T phases, which can be detected experimentally. 
Besides NiS$_2$, there are more achiral non-collinear altermagnet candidates, such as MnTe$_2$~\cite{Zhu2024}, Mn$_3$Ir~\cite{Tomeno_Mn3Ir_1999}, SmCrO$_3$~\cite{Tripathi_SmCrO3_2017}, CoCrO$_4$~\cite{Pernet_CoCrO4_1969,ssg_coplanar-even-wave}, Mn$_3$Cu$_{0.5}$Ge$_{0.5}$N~\cite{Iikubo_Mn3CuGeN_2008}, Mn$_2$GeO$_4$~\cite{White_Mn2GeO4_2012}, etc, whose magnetic atoms all host inversion on-site symmetry. We expect the quadrupolar spin texture and phenomena such as the spin Hall and piezomagnetic effects in those materials. 
There are spin-degenerated achiral cases when the magnetic phase transition enlarges the unit cell and introduces the $\cal{T}$ times lattice fractional translation symmetry, such as YCr(BO$_3$)$_2$~\cite{Sinclair_RCr(BO3)2_2017}, Cr$_2$As~\cite{Yamaguchi_Cr2As_1972}, CrN~\cite{Corliss_CrN_1960}, BiMn$_3$Cr$_4$O$_{12}$~\cite{Zhou_BiMn3Cr4O12_2017}, LaMn$_3$V$_4$O$_{12}$~\cite{Saito_LaMn3V4O12_2014}. The low-T phase of NiS$_2$ also doubles the unit cell, while the non-collinearity preserves only mirror instead of the $\cal{T}$ times lattice translation symmetry, which preserves the altermagnetic nature. 
From our results it is clear that non-collinear altermagnetism drives phenomena such as spin Hall and piezomagnetic effects, as well as spin textures, that are equivalent to having an effective SOC present, even if relativistic SOC are lacking
.
Our work provides a concrete example of altermagnetic phases with multiple magnetic transitions and highlights the rich physics arising from the interplay between symmetry, magnetism in these systems.

\clearpage

\section{Computational methods}

The first-principles calculations were performed using VASP \cite{vasp}, employing the projector augmented wave (PAW) method \cite{paw}. 
The Brillouin zone was sampled on a $7\times7\times7$ and $9\times9\times9$ $k$-point grid centered on the Gamma point for high-T cubic and low-T rhombohedral lattice, respectively. 
The energy cutoff for the plane wave basis was set to 550 eV. 
The Becke-Johnson (mBJ) approximation\cite{mbj_1,mbj_2} for the exchange-correlation functional is adapted for a better band-gap and local magnetic moments consistency of Ni in high-T phase with mBJ = 1.15.  
The Hubbard term was introduced for the low-T phase with a value of 3.0 eV for the $d$ orbitals of the Ni atoms within the DFT+U framework to account for electron-electron correlations. 
The Wannier-based Hamiltonian was symmetrized based on the maximally localized Wannier functions generated by the WANNIER90 interface \cite{wannier}. The projectors were the $d$ orbitals of Ni and $p$ orbitals of S. The magnetic moments within the self-consistent noncollinear ground state.

\noindent{\bf Acknowledgements} \\
We acknowledge fruitful discussions to Fabio Orlandi.  M.G.V. thanks support to the Spanish Ministerio de Ciencia e Innovación (PID2022-142008NB-I00),  the Canada Excellence Research Chairs Program for Topological Quantum Matter and funding from the IKUR Strategy under the collaboration agreement between Ikerbasque Foundation and DIPC on behalf of the Department of Education of the Basque Government.
M.H. thanks Paolo G. Radaelli for enlightening diccussion and Ulrike Nitzsche for technical assistance. 
M.H. thanks the support from the Alexander von Humboldt Foundation and Leibniz Competition Project No. J200/2024.
We acknowledge financial support by the Deutsche Forschungsgemeinschaft (DFG, German Research Foundation), through SFB 1143 (Project ID 247310070), project A05, Project No. 465000489, and the W\"urzburg-Dresden Cluster of Excellence on Complexity and Topology in Quantum Matter, ct.qmat (EXC 2147, Project ID 390858490).\\

\bibliography{nis2.bib}

\newpage

\section*{Supplementary Information}

\renewcommand{\thefigure}{S\arabic{figure}}
\setcounter{figure}{0}
\renewcommand{\thetable}{S\arabic{table}}
\setcounter{table}{0}

{\bf 1. Surface states in both phases}

\begin{figure}[h!]
    \centering
    \includegraphics[width=0.8\linewidth]{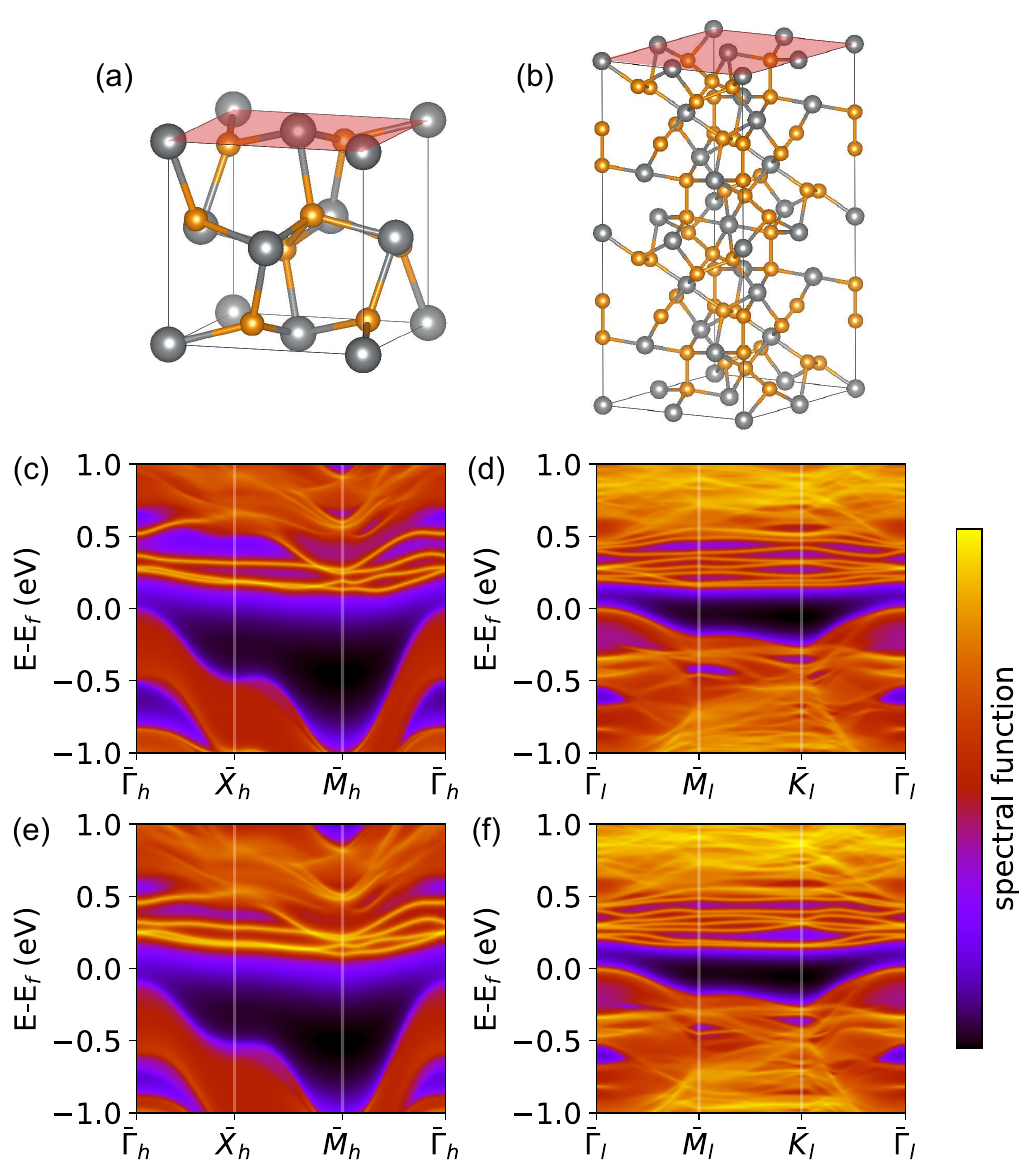}
    \caption{Surface spectrum of NiS$_2$ in (a) the high-T phase and (b) the low-T phase. With the terminations denoted in (a) and (b), the surface states are shown along high symmetry lines without SOC in (c,d) and with SOC in (e,f).}
    \label{fig:surface_states}
\end{figure}

\newpage

{\bf 2. GGA+U test for low-T phase}

The local magnetic moments of Ni in low-T phase are calcualted with varied GGA+U values from 0 to 5 eV. The noncollinear magnetic arrangement is preserved with only the changes of total magnitudes. The experimental values of magnetic moments are denoted in the figure with the dashed grey lines of $|\mathbf{m}_{Ni_{1-2}}| = 1.195 \mu _B$ and $|\mathbf{m}_{Ni_{3-8}}| = 1.208 \mu _B$. Compared with the experimental values, the GGA+U = 3 eV fits well and is used in the main text.
\begin{figure}[h!]
    \centering
    \includegraphics[width=0.8\linewidth]{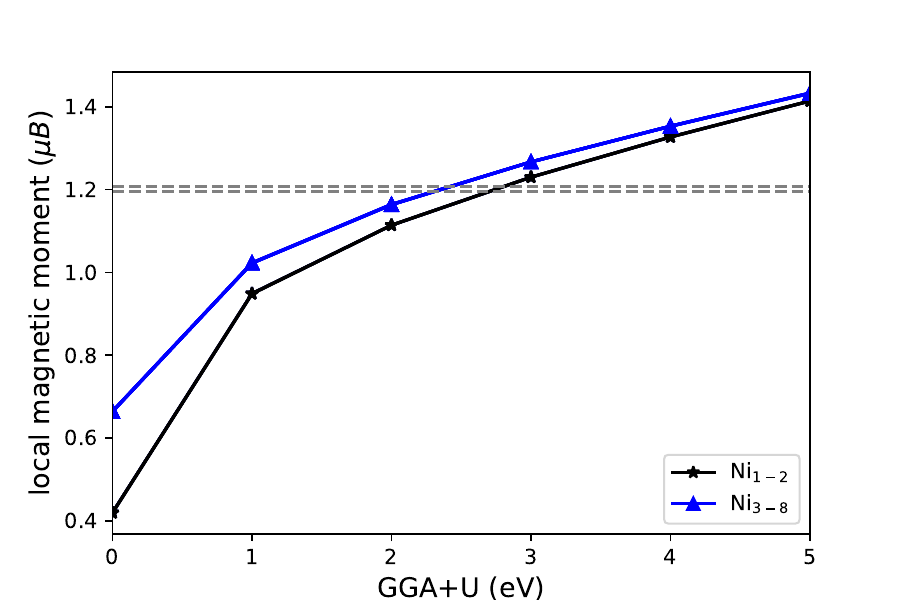}
    \caption{The local magnetic moments of Ni in low-T phase with varied GGA+U values. The experimental values of magnetic moments are denoted in the figure with the dashed grey lines of $|\mathbf{m}_{Ni_{1-2}}| = 1.195 \mu _B$ and $|\mathbf{m}_{Ni_{3-8}}| = 1.208 \mu _B$.}
    \label{fig:GGA+U}
\end{figure}

\newpage

{\bf 3. Band degeneracies in low-T phase}

All the possible IRs at different momenta can be checked\cite{SSG_irreps_2}. We compare the IRs of MSG and SSG for the maximal momenta that the case in the material is needed for further identifying by EBR.

\begin{table}[h!]
    \centering
    \begin{tabular}{c|c|c}
       $\bm{k}$/G  & MSG & SSG \\
       \hline
        $\Gamma:(0,0,0)$ & $\overline{\Gamma_i} (1)$ (i=4 $\sim$ 9) & $\overline{\Gamma_i} (2),\overline{\Gamma_j} (1)$ (i,j=1,2) \\
        \hline
        F : ($\frac{1}{2},\frac{1}{2},0$) & $\overline{F_i} (1)$ (i=2,3) & $\overline{F_i} (1)$ (i = 1,2) \\
        \hline
        L : (0,$\frac{1}{2}$,0) & $\overline{L_i} (1)$ (i=2,3) & $\overline{L_i} (2)$\\
        \hline
        T : ($\frac{1}{2},\frac{1}{2},\frac{1}{2}$) & $\overline{T_i} (1)$ (i=4 $\sim$ 9) & $\overline{T_i} (2)$ (i = 1,2,3)\\
    \end{tabular}
    \caption{The IRs of high symmetry points in low-T phase of NiS$_2$. The Irreps and its dimensions of MSG and SSG are provided, which have big differences at $\Gamma, L,T$.}
    \label{tab:my_label}
\end{table}

{\bf 4. Detailed spin-texture results}

\begin{figure}[h!]
    \centering
    \includegraphics[width=0.9\linewidth]{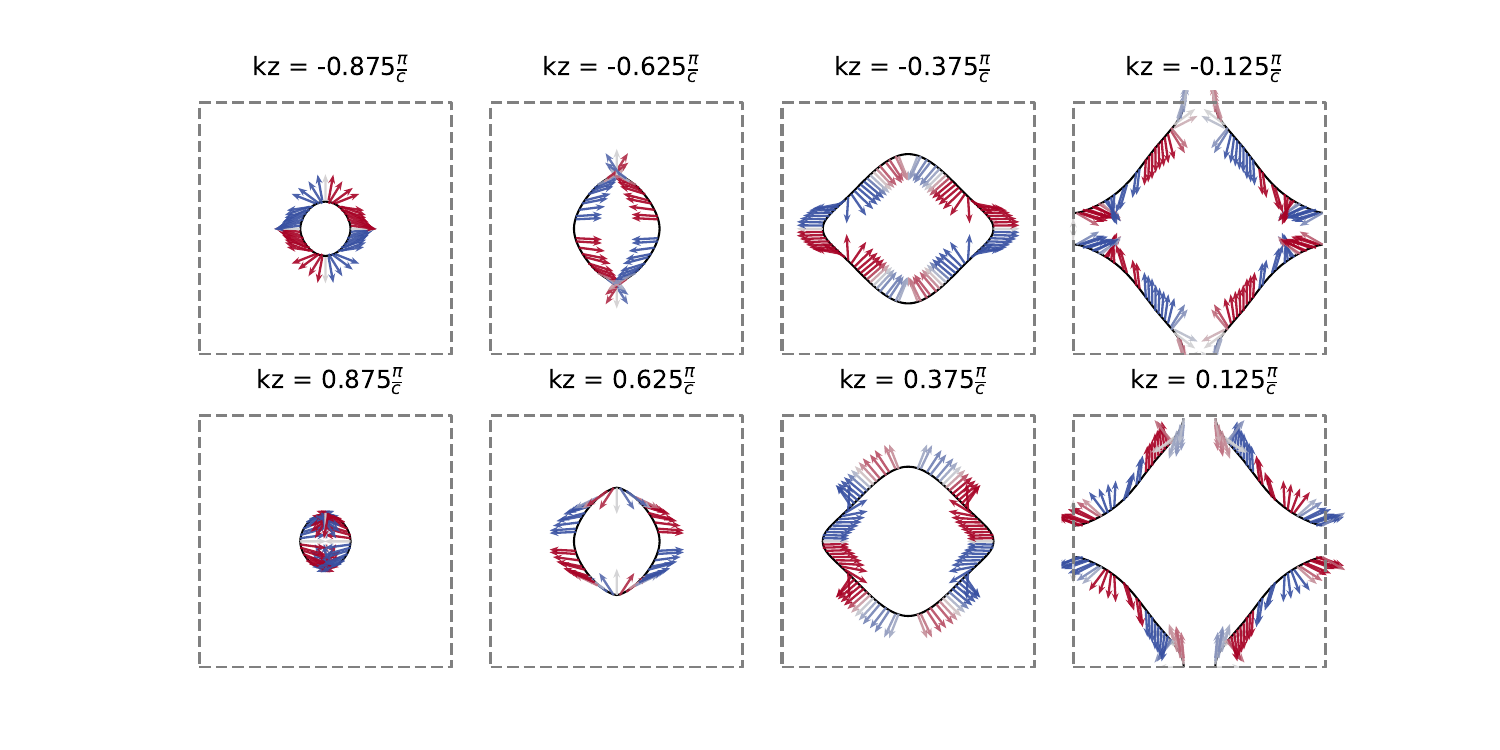}
    \caption{The spin texture at different $k_z$ planes in high-T phase. The color of arrows represents the out-of-plane component $s_z$. The grey dashed lines denote the first Brillouin zone.}
    \label{fig:high-T_ST}
\end{figure}

\begin{figure}
    \centering
    \includegraphics[width=0.75\linewidth]{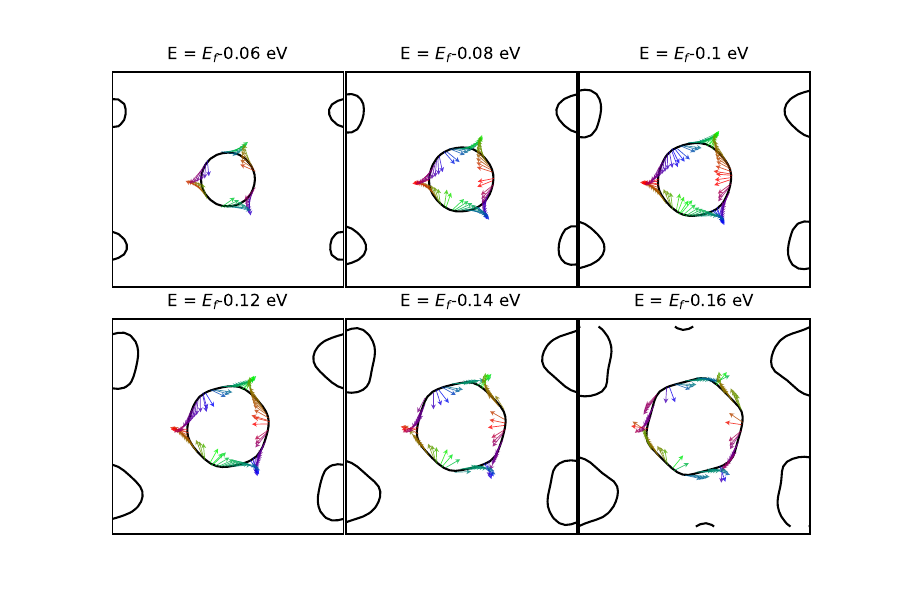}
    \caption{The spin texture at $k_z = 0$ plane with varied energy contours in low-T phase. The color of arrows represents the in-plane $s_x-s_y$ angle with the same colorbar in the main-text.}
    \label{fig:low-T_ST}
\end{figure}

Besides the spin texture results shown in the main text, the spin texture at different $k_z$ planes in high-T phase and with varied energy contours in low-T phase are presented in Fig.~\ref{fig:high-T_ST} and Fig.~\ref{fig:low-T_ST}, respectively. All the analyses in the main text are consistent with these supplementary results, e.g., the quadrupolar pattern and in-plane spin flips between $\pm k_z$ planes in high-T phase, and the coplanar spin texture in low-T phase.

\end{document}